\documentclass[preprint,showpacs,preprintnumbers,amsmath,amssymb]{revtex4}
\usepackage{graphicx}% Include figure files
\usepackage{dcolumn}% Align table columns on decimal point
\usepackage{bm}% bold math

\begin{document}
\title{Unwinding of circular helicoidal molecules versus size }

\author{Marco Zoli}

\affiliation{School of Science and Technology - CNISM \\  University of Camerino, I-62032 Camerino, Italy \\ marco.zoli@unicam.it}

\date{\today}

\begin{abstract}
The thermodynamical stability of a set of circular double helical molecules is analyzed by path integral techniques. The minicircles differ only in \textit{i)} the radius and \textit{ii)} the number of base pairs ($N$) arranged along the molecule axis. Instead, the rise distance is kept constant. For any molecule size, the computational method simulates a broad ensemble of possible helicoidal configurations while the partition function is a sum over the path trajectories describing the base pair fluctuational states. The stablest helical repeat of every minicircle is determined by free energy minimization. We find that, for molecules with $N$ larger than $100$, the helical repeat grows linearly with the size and the twist number is constant. On the other hand, by reducing the size below $100$ base pairs, the double helices sharply unwind and the twist number drops to one for $N=\,20$. This is predicted as the minimum size for the existence of helicoidal molecules in the closed form. The helix unwinding appears as a strategy to release the bending stress associated to the circularization of the molecules.

\end{abstract}


\maketitle

\section*{ Introduction }

A large amount of physical and biochemical research on circular DNA has been carried out over the last fifty years following the first observations that the single stranded DNA of coliphage $\phi$X174 was a ring
\cite{fiers} and the tumor virus DNA was in the form of covalently closed circular duplex molecules \cite{dulbecco,vinograd}.
Being less degradable than linear DNA, circular DNA  has recently revealed a potential for those biotechnological applications such as assembly  of nanotubes which require both rigidity and flexibility of the building blocks
\cite{willner,xiao}.

Bending flexibility of short DNA fragments has been appreciated since it was pointed out \cite{cloutier} that linear molecules  of order $100$ base pairs (\textit{bps}) can spontaneously close into circles, even in the absence of proteins, with much higher cyclization probabilities than those predicted by conventional worm like chain models \cite{shore,shimada,vafa}.
The circularization of a DNA fragment causes a bending stress in the loop that may be released by a local untwisting of the double helix with formation of fluctuational bubbles comprising a few transiently broken \textit{bps} \cite{yan04}. While bubbles are essential to key biological functions such as replication and transcription of the genetic code \cite{benham,hwa,lavery,kalos,rapti},  their length is related to the degree of supercoiling as shown by atomic force microscopy imaging of circular plasmids  \cite{metz12}.

In general, bending and twisting degrees of freedom are intertwined in DNA  and their interplay crucially depends on the molecule size.

Recently, short sequences of extra-chromosomal circular DNA have been found, both in mouse tissues and human cells, in a range of $80-2000$ \textit{bps} with length distributions peaked between $200$ and $400$ \textit{bps} \cite{dutta}. These findings lend support to the prediction that there should be a minimum size for the existence of double stranded loops whereas, below that size, circular molecules can be only single stranded \cite{olsen}. However, the determination of such threshold is still an open question also in view of the fact that duplex DNA minicircles have been obtained, by a modified ligase-assisted protocol \cite{volo08}, even in the $60$ \textit{bps} range.

Specifically in the latter, disruptions of the regular helical structure have been detected by single-strand-specific endonuclease experiments carried out on DNA minicircles with various radii \cite{volo08}. Such local disruptions may appear either in the form of kinks which move one base pair out of the stack \cite{volo05,maddocks} or as a breaking of a few base pair hydrogen bonds \cite{zocchi13}, the latter event being energetically more costly.

These issues have been attacked by a theoretical model based on the path integral formalism \cite{io09}
and applied to three minicircles  having comparable relative contents of AT- and GC-\textit{bps}, albeit with different radii \cite{io14a}: analysis of the bubble profiles has shown that base pair disruptions and helix unwinding are in fact more pronounced in the smallest minicircle with $66$ \textit{bps} (corresponding to a bending angle of $\sim 6^o$) while larger loops are relatively stabler. Whereas these findings confirm the soundness of the path integral method, it is still unsettled whether a critical radius of curvature for the appearance of  \textit{bps} disruptions in small circles indeed exists and to which extent such critical value may depend on the DNA sequence specificities.

To shed light on these issues we investigate in this work the thermodynamical stability of a broader set of molecules spanning the range between 20 and 140 \textit{bps}. In view of the shortness of these systems we assume that, for any given length, the cyclization occurs with finite probability and the molecules are thus in the circular form \cite{crothers2}. Furthermore, we admit that the minicircles could be found either in the single or in the double stranded helicoidal configuration, keeping the model general enough to deal with both cases. 
For each minicircle size, we simulate a large ensemble of possible twist configurations and compute the free energy profiles as a function of the average helical repeat, i.e. the number of \textit{bps} per helix turn. This methods permits to determine the energetically most convenient helicoidal conformation for a molecule that, by virtue of the circularization process, has incorporated a certain amount of bending stress. 

\section*{ Model }

In general, the quantum statistical partition function of a system is determined by performing an analytic continuation of the quantum mechanical partition function to the imaginary time $\tau=\,it$, $t$ being the real time \cite{fehi}.   $\tau$ is defined within a range whose upper bound is given by the inverse temperature as in the Matsubara method \cite{matsu}. Accordingly, the Euclidean action takes the place of the mechanical canonical action and, in the path integral formalism \cite{feyn}, the statistical partition function is an integral in the configuration space over (closed) paths running along an imaginary time axis. The largest contribution to the statistical partition function comes from those trajectories for which the sum of the kinetic and potential energy is small. The classical partition function is obtained by replacing the quantum thermal wavelength with the classical one as described in detail in \cite{io14}.

The focus of our investigation is a system made of
a homogeneous set of $N$ free purine-pyrimidine \textit{bps} with reduced mass $\mu$. Let's define $\textbf{r}_i$ as the inter-strand fluctuational vector describing the dynamics of the stretching mode for the $i-th$ base pair. 
The application of the path integral method to our classical system relies on the idea that, at finite temperature, the base pair displacements can be thought of as trajectories, $x_i(\tau)$:

\begin{eqnarray}
\pm |\textbf{r}_i| \rightarrow  x_i(\tau) ; \, \, \, \, \tau \in [0 \,, \beta ] \,,
\label{eq:005}
\end{eqnarray}

where $\beta=\,(k_B T )^{-1}$, $k_B$ is the Boltzmann constant and $T$ is the temperature.  
As the trajectories are closed, $x_i(0)=\, x_i(\beta)$, they can be expanded in Fourier series: 

\begin{eqnarray}
& &x_i(\tau)=\, (x_0)_i + \sum_{m=1}^{\infty}\Bigl[(a_m)_i \cos(\omega_m \tau ) + (b_m)_i \sin(\omega_m \tau ) \Bigr] \, \nonumber
\\
& &\omega_m =\, \frac{2 m \pi}{\beta} \, .
\label{eq:01}
\end{eqnarray}

The set of coefficients $\{(x_0)_i, (a_m)_i, (b_m)_i \}$ represents a point in the path configuration space corresponding to a fluctuational state for the $i-th$ base pair.
Consistently one defines, over the space of the Fourier coefficients,  an integration measure $\oint {D}x_i$ which normalizes the partition function $Z_{k}$ for the freely fluctuating $N$ particles:

\begin{eqnarray}
& &\oint {D}x_i \equiv {\frac{1}{\sqrt{2}\lambda_C}} \int d(x_0)_i \prod_{m=1}^{\infty}\Bigl( \frac{m \pi}{\lambda_C} \Bigr)^2 \int d(a_m)_i \int d(b_m)_i \, \, \nonumber
\\
& & Z_{k}=\, \prod_{i=1}^{N} \oint Dx_i \exp \bigl[- A_{k}[x] \bigr] =\, 1 \, \nonumber
\\
& &A_{k}[x]=\, \sum_{i=1}^{N} \int_{0}^{\beta} d\tau \biggl[ \frac{\mu}{2} \dot{x}_i^2(\tau) \biggr] \, .
\label{eq:001}
\end{eqnarray}

$\lambda_C$ is the classical thermal wavelength and $A_{k}[x]$ is the kinetic action which, as a consequence of the real space mapping, is a dimensionless quantity.
Hence, Eq.~(\ref{eq:001}) sets the zero for the free energy, $\beta^{-1} \ln Z_{k}$,  of the system.

\begin{figure}
\includegraphics[height=9.0cm,width=9.5cm,angle=-90]{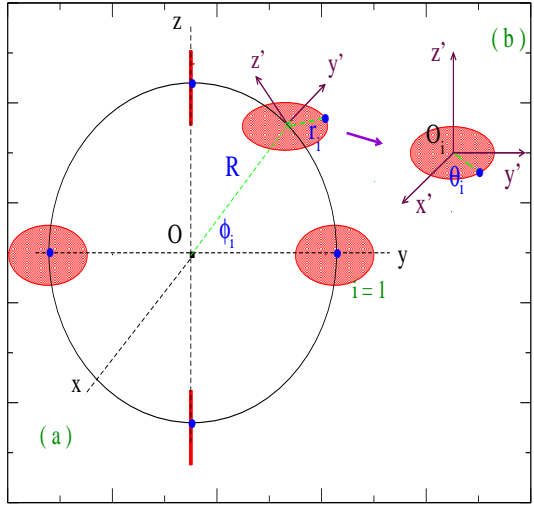}
\caption{\label{fig:1}(Color online)  
Schematic of the base pairs (blue dots) stacked along the circular molecule backbone whose radius $R$ lies on the $(y,\, z)$ plane. $\textbf{r}_{i}$ is the inter-strand fluctuational vector, for the $i-th$ base pair, spanning the red shaded orbit. 
For $|\textbf{r}_{i}|=\,0$, the orbit shrinks into the fluctuations free state. Adjacent orbits along the molecule axis are bent by an angle $\phi_i$. $\theta_i$ is the twist angle for the $i-th$ base pair as shown in the reference system $\textbf{O}_i$.
 }
\end{figure}

Next, as shown in Fig.~\ref{fig:1}, we uniformly arrange the $N$ objects (represented by the blue points) on a circle, with radius $R$, which represents the bent molecule axis.  The $i-th$ object can fluctuate with respect to $R$ and its vector ($\textbf{r}_i$) describes a (red shaded) orbit lying on a plane which is bent by $\phi_i$  with respect to the $(x,\,y)$ plane.  In the local reference system, centered in $\textbf{O}_i$, the $i-th$ object makes a positive twist angle ($\theta_i$) with respect to the $\textbf{x}^{'}$ axis and adjacent objects along the stack are twisted by a constant angle. 
The rise distance between neighbor objects along the circumference is: $d=\,2 \pi R / N$. Thus we have built a circular model for a set of particles arranged in a helicoidal conformation. The model can be applied to both a single stranded and a double stranded molecule. In the former case, $\textbf{r}_i$ represents the $i-th$ base along the strand. In the latter, $\textbf{r}_i$ is the inter-strand fluctuational vector, for the $i-th$ base pair, with respect to the equilibrium position which corresponds to the minimum for the hydrogen bond potential (see below). In the following we will refer to a set of $N$ \emph{bps} with the caveat that very short circular molecules may more likely be found in the single strand configuration. The values $d=\,3.4$\AA  \, and $\mu=\,300 \,amu$ peculiar of DNA molecules are hereafter taken.

Accordingly, with respect to the \textbf{O} reference system (see Fig.~\ref{fig:1}), the general vector  $\textbf{t}_i$ for the $i-th$ base pair is:

\begin{eqnarray}
& &\bigl({t }_i \bigr)_{x} =\, |\textbf{r}_i| \cos\phi_i \cos\theta_i \, \nonumber
\\
& &\bigl({t }_i\bigr)_{y} =\,(R + |\textbf{r}_i|\sin\theta_i) \cos\phi_i
\, \nonumber
\\
& &\bigl({t }_i\bigr)_{z} =\,(R + |\textbf{r}_i|) \sin\phi_i \, .
\,
\label{eq:004}
\end{eqnarray}

The $x$-axis is normal to the sheet plane and $R$ lies on the $(y,\,z)$ plane.
For small size molecules, it can be reasonably assumed that the relative bending between adjacent orbital planes along the stack is constant and depends only on the circle length. Then, the bending angle is, $\phi_i =\, (i-1){{2 \pi} / N} + \phi_S$,    
while the twisting is measured by, $\theta_i =\, (i - 1) 2\pi / h + \theta_S$, where $h$ is the number of \textit{bps} per helix turn \cite{depew,cozza}. 
The computation sums over a distribution of $\phi_S$ and $\theta_S$ in order to avoid to pin the first base pair in the sequence to a specific angular position.
Furthermore, as the molecule axis lies on a plane, the Writhe (measuring the spatial coiling of the axis itself \cite{ful1} ) vanishes and the integer Linking number ($Lk$) for the closed-circular molecule coincides with the Twist number $Tw$ \cite{bates,io14b}. The latter is given by $Tw=\, N / h$ which, however, is generally not an integer. While in long molecules one can take $Lk$ as the closest integer to $N / h$, the extra twist required to join the strand ends becomes large in short sequences thus significantly reducing the cyclization probability \cite{shore}. Accordingly, as $h$ is here a variable, we assume to deal with closed-circular molecules only if  $N / h$ is an integer and, in these cases, also closure conditions on the radial fluctuations are implemented in the computation, $ |\textbf{r}_{N+1}| =\, |\textbf{r}_{1}| $. Whenever such conditions are not fulfilled, our circular molecules are open.

As the efficacy of the method crucially depends on the partition of the $Tw$ range, we take a small incremental step, $\Delta Tw =\,0.0125$, so that any molecule may assume in principle a large ensemble of possible twisted conformations, each with a specific $h$. The computational task consists in evaluating the free energy profiles within a broad range of $h$ values. It follows that the CPU time grows fast with the size of the molecule.

In analogy with Eq.~(\ref{eq:005}), the path integral method is applied to the bent configuration. Accordingly, $\textbf{t}_i$ maps onto a time dependent path fluctuation:

\begin{eqnarray}
& & |\textbf{t}_i| \rightarrow  \eta_i(\tau)\, , \,
\label{eq:005a}
\end{eqnarray}

which, in the circular geometry, reads:

\begin{eqnarray}
& & \eta_i(\tau)=\, \bigl[R^2 + x_i(\tau)^2 + 2R |x_i(\tau)|f(\theta_i,\phi_i) \bigr]^{1/2} \, \nonumber
\\
& &f(\theta_i,\phi_i)=\,\sin\theta_i \cos^2 \phi_i + \sin^2 \phi_i  \,.
\label{eq:005b}
\end{eqnarray}

Then, the partition function for the ensemble of $N$ \textit{bps}  oscillating around the fluctuations free circle is given by:

\begin{eqnarray}
& & Z_{0}=\, \prod_{i=1}^{N} \sum_{\theta_S,\, \phi_S } \oint Dx_i  \exp \bigl[- A_{0}[\eta ] \bigr] \, \nonumber
\\
& &A_{0}[\eta ]=\, \sum_{i=1}^{N} A_0[\eta_i] \, \nonumber
\\
& &A_0[\eta_i]\equiv \,  \int_{0}^{\beta} d\tau \biggl[ \frac{\mu}{2} \dot{\eta }_i^2(\tau) \biggr] \, . \,
\label{eq:0001}
\end{eqnarray}

The action $A_{0}[\eta ]$ accounts for the bending energy cost due to the deformation which closes the array of $N$ free particles into a loop. This is proved by the following analytical argument.

Take for simplicity only one Fourier component in  Eq.~(\ref{eq:01}) (and set  $(b_1)_i=\,0$). Then the path amplitude can be approximated by: $x_i(\tau)\sim \, (x_0)_i + (a_1)_i \cos(\omega_1 \tau )$ and
$A_0[\eta_i]$ in Eq.~(\ref{eq:0001}) reads:

\begin{eqnarray}
& &\int_{0}^{\beta} d\tau \biggl[ \frac{\mu}{2} \dot{\eta }_i^2(\tau) \biggr] \approx \, \, \frac{2 (s_1)^2_i}{\beta } \int_{0}^{\beta} d\tau \sin^2 \biggl(\frac{2\pi \tau }{\beta }\biggr)g_i(\tau) \, \nonumber
\\
& &g_i(\tau)\equiv \, \frac{\bigl[|x_i(\tau)| + R f(\theta_i,\phi_i) \bigr]^{2}}{\bigl[R^2 + x_i(\tau)^2 + 2R |x_i(\tau)|f(\theta_i,\phi_i) \bigr]} \, \, \nonumber
\\
& &(s_1)^2_i \equiv \frac{\pi^3 (a_1)^2_i}{\lambda^2_C} \, \, . 
\label{eq:2}
\end{eqnarray}

As the action is not quadratic in the Fourier coefficients,  $Z_{0}$ does not decouple into a product of Gaussian integrals. This marks the difference with respect to $Z_{k}$ in Eq.~(\ref{eq:001}).
Setting $R=\,0$,  the Gaussian approximation is recovered for any Fourier component, $A_0[\eta_i] \approx  \, (s_1)^2_i$ and $Z_0 \rightarrow Z_k$. 

In general, for a circular molecule with finite $R$, the normalization condition in Eq.~(\ref{eq:001}) is not fulfilled. Hence, the free energy level for the loop, $F_0=\,\beta^{-1} \ln Z_{0}$, is not zero. 
$F_0$ is computed via Eq.~(\ref{eq:0001}) with the integration measure given in the first of Eqs.~(\ref{eq:001}).

A mesoscopic model for double stranded DNA molecules should incorporate in the action also the terms describing the effective hydrogen bond base pair interactions.  This is done by including: \textit{a)} the Morse potential, $\,V_{M}[ \eta _i(\tau)]=\, D_i \bigl[\exp(-b_i (\eta_i(\tau) - R)) - 1 \bigr]^2$, which also accounts for the repulsive electrostatic interaction between phosphate groups. 
$D_i$ is the pair dissociation energy while the inverse length $b_i$ sets the potential range. \textit{ b)} A solvent potential which enhances the barrier for base pair breaking and stabilizes the strands: $V_{Sol}[\eta_i(\tau)]=\, - D_i f_s \bigl(\tanh((\eta_i(\tau) - R)/ l_s) - 1 \bigr)$  \cite{collins}. The factor $f_s$ is related to the counterions concentration in the solvent and $l_s$ tunes the width of the solvent barrier. Both terms and parameters choice,   $D_{i}=\,45\, meV$,  $b_{i}=\,2$ \AA$^{-1}$, $f_s=\,0.1$, $l_s=\,0.5$ \AA, are described in detail in \cite{io14b}.  Hence the partition function for our homogeneous \textit{ds}-circular molecules reads:

\begin{eqnarray}
& & Z_{1}=\, \prod_{i=1}^{N} \sum_{\theta_S, \phi_S} \oint Dx_{i}  \exp \bigl[- A_{1}[\eta ] \bigr] \, \nonumber
\\
& &A_{1}[\eta ]=\, \sum_{i=1}^{N} \int_{0}^{\beta} d\tau \biggl[ \frac{\mu}{2} \dot{\eta }_i^2(\tau) + V_{1}[ \eta _i(\tau)]  \biggr] \, \nonumber
\\
& &V_{1}[ \eta _i(\tau)]=\, V_{M}[ \eta _i(\tau)] + V_{Sol}[\eta_i(\tau)]\, . 
\label{eq:0002a}
\end{eqnarray}

and the free energy is computed via \, $F_1=\,\beta^{-1} \ln Z_{1}$ \cite{note}. 

\section*{Results and Discussion}

Given a molecule with $N$ \emph{bps}, for any twist configuration, the programme sums over the sets of Fourier coefficients corresponding to $\sim 10^5$ fluctuational states for each base pair. This suffices to get numerical convergence in the partition function. The computation includes also large (of the order of a few \AA ngstroms) path amplitudes $x_i(\tau)$ which measure the distance between the pair mates on complementary strands. However, such amplitudes should not exceed the radius $R$ in order to preserve the overall circular shape of the molecule. Technically, too large fluctuational amplitudes can be discarded as they would yield a high kinetic action (see Eq.~(\ref{eq:2})) hence, a small contribution to the partition function.

\begin{figure}
\includegraphics[height=7.1cm,width=9.5cm,angle=-90]{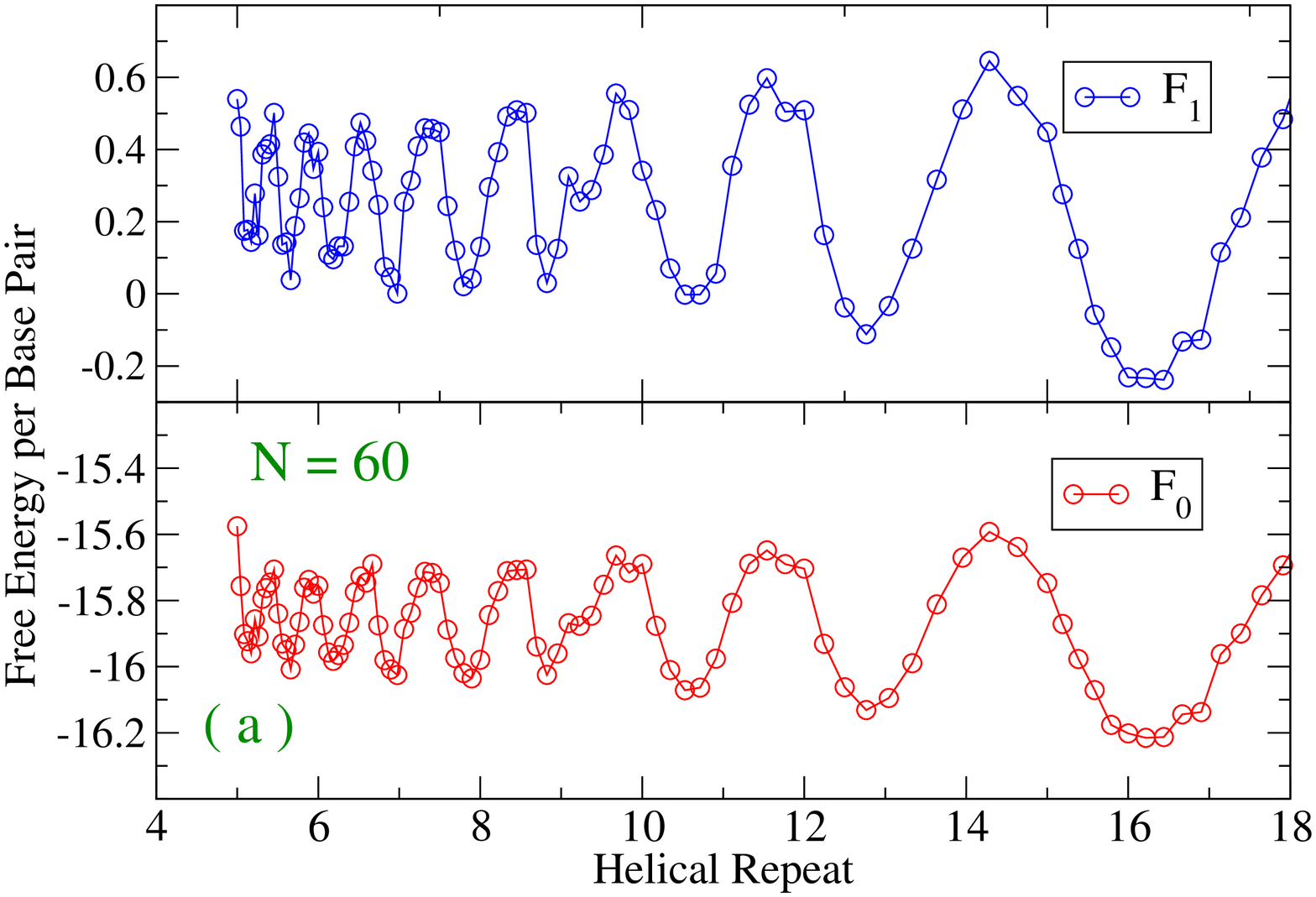}
\includegraphics[height=7.1cm,width=9.5cm,angle=-90]{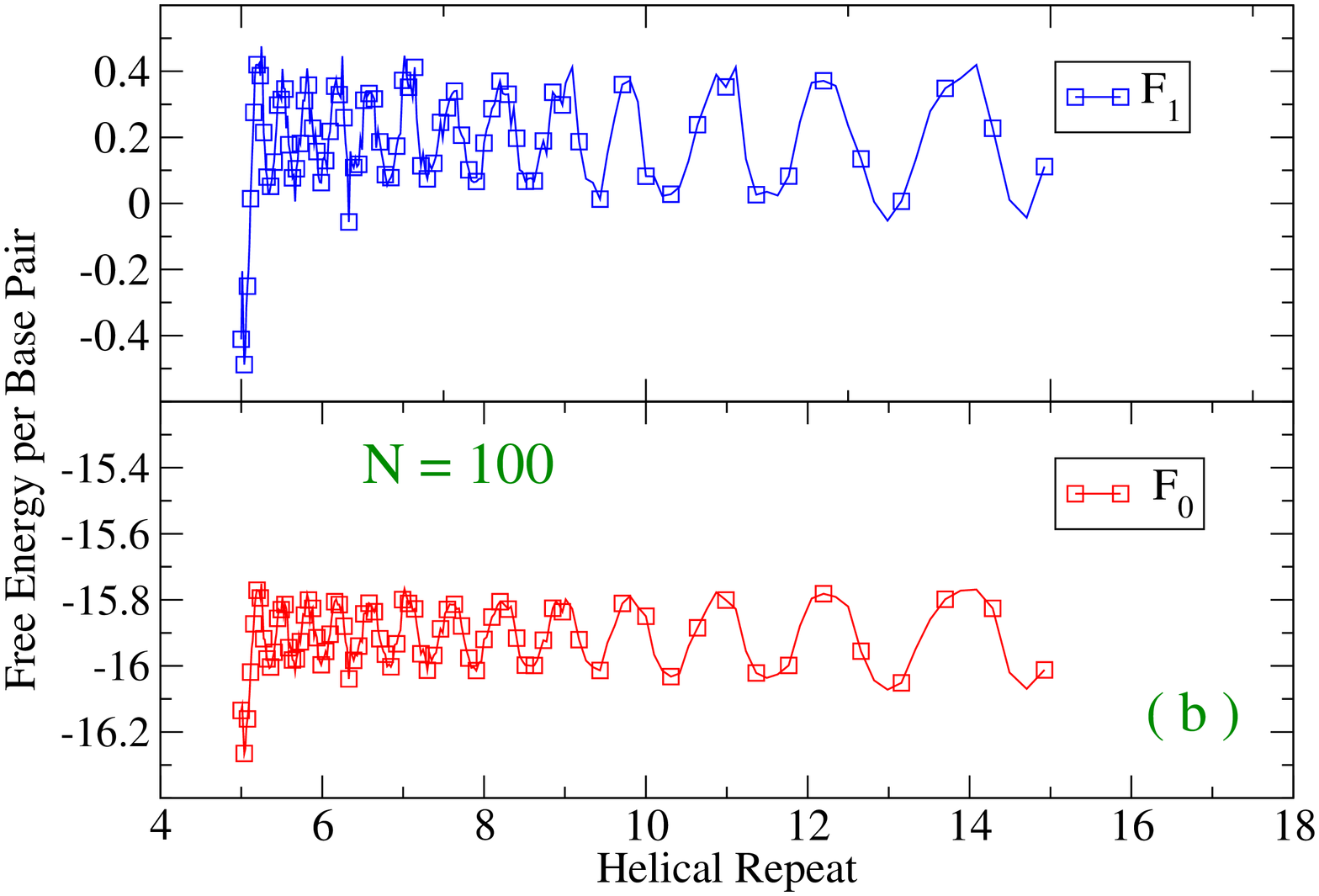}
\includegraphics[height=7.1cm,width=9.5cm,angle=-90]{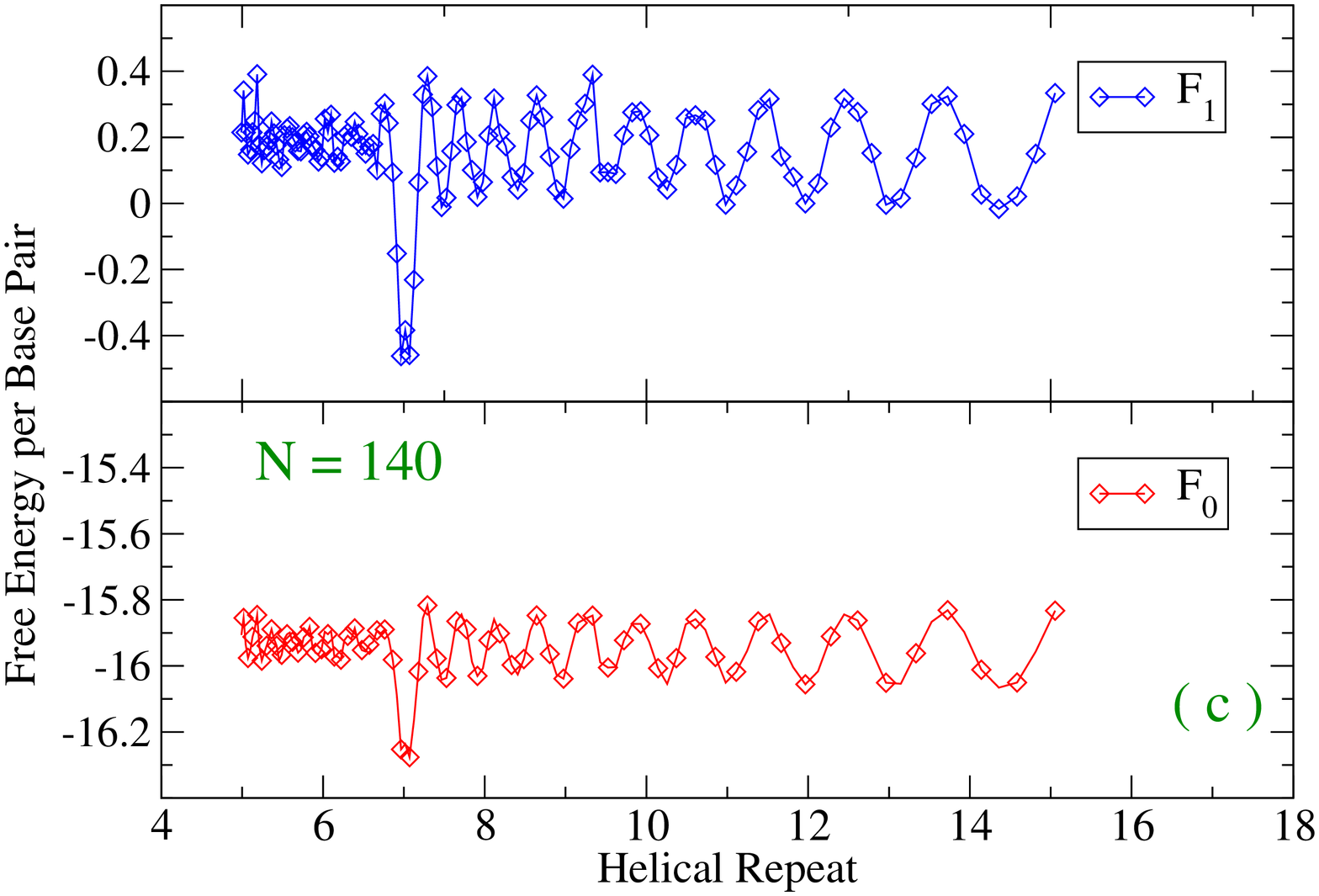}
\caption{\label{fig:2}(Color online)  
Room temperature free energies (in meV) for three helicoidal circular molecules as function of the helical repeat. $F_0$ and $F_1$ are computed by Eqs.~(\ref{eq:0001})
and Eqs.~(\ref{eq:0002a}), respectively.}
\end{figure}

The free energy profiles are shown in Figs.~\ref{fig:2} for three circular molecules. $F_0$ and $F_1$ are compared in each panel.
The calculations are carried out at room temperature being aware that thermal effects would loosen the base pair bonds, favor the molecules unwinding, ultimately leading to denaturation in the high temperature regime \cite{santos}.
The input parameters of the model, i.e. rise distance, reduced mass and hydrogen bond potential, are common to all circles.

While oscillating patterns are found in all the \textit{free energy versus $h$} plots, the location of the free energy minimum strongly varies with the molecule size. This indicates that, for each minicircle, the thermodynamically stablest configuration is associated to a specific twist conformation.
In particular, the smallest circle (among the three ones displayed) takes the equilibrium state with the largest helical repeat, $h \sim 16$, suggesting that the helix unwinding is more pronounced in small size molecules where the  bending of the backbone is strong. 
Importantly, in each panel, the minima for $F_1$ coincide with those for $F_0$  signaling that the molecule unwinding is driven by the bending rather than by the hydrogen bonds, at least at room temperature. Then, the unwinding patterns of single and double stranded circular molecules are similar.  On the other hand, the potential in Eqs.~(\ref{eq:0002a}) contributes to order the molecular structure and, accordingly, the  $F_1$ values are shifted upwards on the energy scale with respect to $F_0$.

\begin{figure}
\includegraphics[height=9.0cm,width=9.5cm,angle=-90]{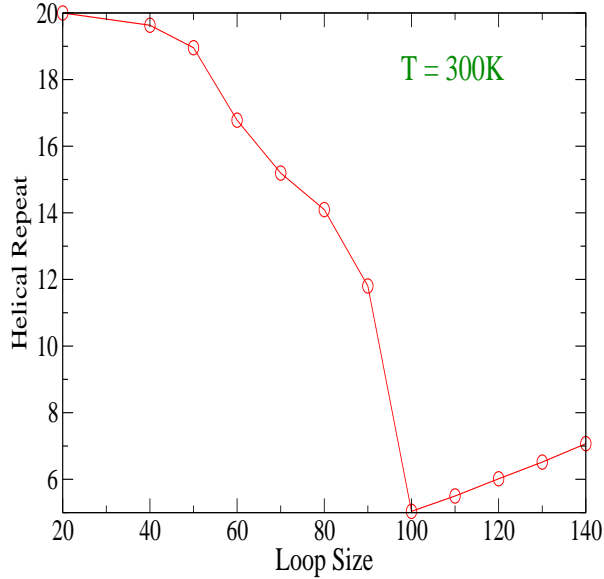}
\caption{\label{fig:3}(Color online)  
Helical repeat values obtained by minimization of the free energy of minicircles with different size.}
\end{figure}

Fig.~\ref{fig:3} summarizes the results of the free energy computation for a set of circles with size $N \in [20,\, 140]$. The helical repeat values which minimize the free energies are reported. Interestingly the curve is not monotonous and the most twisted conformation ($h \simeq \,5$) is found for $N=\,100$.
For  $100 < N < 140$, the $h$ equilibrium values increase smoothly and linearly with $N$ so that the twist number remains essentially constant, $Tw =\, 20$. 
This is however a property which should be investigated also for larger circles than those here examined.

Instead, for $N < \,100$, $h$ sharply grows. This is physically consistent with the fact that short molecules, once bent, have small angles between adjacent \textit{bps} along the stack. As a consequence, short circular molecules tend to untwist in order to attain an energetic configuration which may efficiently release the bending stress. Furthermore, $N / h$ is generally not an integer indicating that, in the very small size range, circular molecules are unlikely to be found in the closed conformation.

Only for the loop with $N=\,20$, one single turn of the helix involves all $20$ \textit{bps} that is, $Tw =\,1$ and the bending angle is $18^o$. It follows that, in our model, no closed helicoidal molecules can be found under such size: a prediction which would be interesting to check experimentally.

The helical repeats  selected in Fig.~\ref{fig:3} represent the most probable twist conformations associated to the deepest minimum in the free energy profiles with multiple wells. It should be however remarked that thermal effects may transiently drive every molecule out of its deepest well bringing it to one of the energetically close states associated to other $h$ values \cite{singh}.

In conclusion, the picture emerging from this path integral computation points to the
existence of a critical size of order $100$ \textit{bps} below which the molecule helix progressively unwinds. The applied model is based on the actions in Eq.~(\ref{eq:0001}) and Eq.~(\ref{eq:0002a}) accounting for the kinetic energy and the stabilizing hydrogen bond potential.  A more structured description
including also the intra-strand base pair stacking may certainly refine the predictions for the most probable helical repeat values of specific sequences but it is not expected to change the trend of the results for the helical unwinding versus $N$ presented in this work \cite{note1}. 

Thus, applying a criterion of thermodynamical stability, we have found that the unwinding of small helicoidal circular molecules hosting a uniform distribution of nucleotides is  driven by the size of the loop and it is not essentially affected by the sequence specificities. The helix unwinds in order to release the stress associated to the bending of the molecule backbone. 
The equilibrium twist conformations have been determined by minimizing the molecules free energy under the assumption that the bending angle between adjacent base pairs along the stack is constant. For circles larger than those here examined, such assumption may not hold and, accordingly, bending fluctuational effects \cite{menon,mazur} should be incorporated in the theoretical description. Finally, for short double stranded molecules, a direct computation of the cyclization probabilities in the framework of the path integral method may contribute to clarify whether a threshold size for closed circular conformations indeed exists.


\begin{thebibliography}{widest-label}
\bibitem{note}
Note that the fluctuations free configuration, $x_i(\tau)=\,0$, defines the zero for the one particle potential $V_{1}[ \eta _i(\tau)]$. One may also consider the fluctuations with respect to the finite inter-strand equilibrium distance: this would change the fluctuation amplitudes and shift the zero for $V_{1}[ \eta _i(\tau)]$ but it would not change the results which follow from the free energy minimization.
\bibitem{note1}
Base pair stacking can be implemented in the model, see e.g. \cite{io14b}, but, being a two particle potential, this would greatly enhance the computational time. Here I have chosen a much finer (and time costly) partition of the $h$ axis than in \cite{io14b}, dropping the stacking term as it is not essential to the purpose of this work.
\end{thebibliography}
\end{document}